\documentclass[12pt]{article}

\usepackage{graphicx,bbm,amscd}

\newcommand{\boldm}[1]{\mbox{\boldmath ${#1}$}}

\begin{document}

\begin{flushright}
  DESY-03-074 \\
  HD-THEP-03-29 \\
  hep-ph/0306247
\end{flushright}

\vspace{\baselineskip}

\begin{center}
\textbf{\Large Probing triple gauge couplings \\[0.3em]
        with transverse beam polarisation \\[0.3em]
        in \boldm{e^+e^- \rightarrow W^+W^-} \\}
\vspace{4\baselineskip}
{\sc M. Diehl}\footnote{email: markus.diehl@desy.de} \\
\vspace{1\baselineskip}
\textit{Deutsches Elektronen-Synchrotron DESY, 22603 Hamburg, Germany} \\
\vspace{2\baselineskip}
{\sc O. Nachtmann\footnote{email: O.Nachtmann@thphys.uni-heidelberg.de}
and F. Nagel\footnote{email: F.Nagel@thphys.uni-heidelberg.de}} \\
\vspace{1\baselineskip}
\textit{Institut f\"ur Theoretische Physik, Philosophenweg 16, 69120
Heidelberg, Germany} \\
\vspace{2\baselineskip}
\textbf{Abstract}\\
\vspace{1\baselineskip}
\parbox{0.9\textwidth}{The prospects of measuring triple gauge
couplings in $W$ pair production at future linear colliders with
transverse beam polarisation are studied.  We consider $CP$ conserving
and $CP$ violating couplings with both real and imaginary parts.  The
maximum achievable sensitivity in a simultaneous measurement of all
couplings is determined using optimal observables, extending an
earlier analysis for longitudinal beam polarisation.  We find good
sensitivity to the coupling ${\rm Im}(g_1^R + \kappa_R)$, which is not
measurable with longitudinal polarisation.  In contrast, for the real
parts (including the $CP$ violating couplings) the sensitivity is
better if both beams are longitudinally polarised.  We conclude that a
comprehensive measurement of all triple gauge couplings requires both
transverse and longitudinal beam polarisation.}
\end{center}
\vspace{\baselineskip}

\pagebreak


\section{Introduction}
\label{sec-intro}

Future linear $e^+e^-$ colliders like TESLA \cite{Richard:2001qm} or
CLIC \cite{Ellis:1998wx} offer remarkable opportunities to probe the
Standard Model (SM) and its numerous proposed extensions.  The wide
c.m.\ energy range from 90~GeV to 800~GeV or possibly 1~TeV at TESLA
and from 500~GeV to 5~TeV at CLIC, the high integrated luminosities in
the inverse attobarn region, the clean environment of $e^+e^-$
collisions, and the possibility to use polarised beams allow for a
variety of precision measurements of the electroweak interactions.
Here we consider the triple gauge couplings (TGCs) $\gamma WW$ and
$ZWW$.

The TGCs are interesting observables for several reasons: Firstly, the
most general $\gamma WW$ and $ZWW$ vertices contain altogether 14
complex parameters \cite{Hagiwara:1986vm}, six of them $CP$ violating.
In the SM the TGCs are predicted by the non-Abelian gauge symmetry,
and only four $CP$ conserving real couplings are non-zero at tree
level.  A variety of new physics effects can manifest itself by
deviations from the SM predictions \cite{Suzuki:1985yh}.  Secondly, in
reactions where longitudinal $W$ boson states are produced via TGCs
the measurement of these couplings may provide information about the
mechanism of electroweak symmetry breaking \cite{Kilian:2003yw}.
Thirdly, though no deviation from the SM has been found for the TGCs
from LEP data \cite{Heister:2001qt}, the bounds obtained are
comparatively weak.  The tightest bounds on the anomalous couplings,
i.e.\ on the differences between a coupling and its SM value, are of
order $0.05$ for $\Delta g_1^Z$ and $\lambda_{\gamma}$, of order $0.1$
for $\Delta \kappa_{\gamma}$ and of order $0.1$ to $0.6$ for the real
and imaginary parts of $C$ and/or $P$ violating
couplings.\footnote{These numbers correspond to fits where all
anomalous couplings except one are set to zero.}  Moreover, many
couplings, e.g.\ the imaginary parts of $C$ and $P$ conserving
couplings, have been excluded from the analyses so far.

For various measurements at future colliders, longitudinal
polarisation of one or both beams is expected to significantly improve
the sensitivity, see e.g.~\cite{Moortgat-Pick:1999ck}.  A dedicated
study of \mbox{$e^-e^+ \rightarrow W^-W^+$} has been performed in
\cite{Diehl:2002nj}.  Longitudinal beam polarisation indeed enhances
the sensitivity to most TGCs.  The linear combination of couplings
\mbox{${\rm Im}(g_1^R + \kappa_R)$} can however not be measured from
the normalised event distribution, unless the beam polarisation is
\emph{transverse}.  Further information on this coupling is contained
in the total event rate as discussed in \cite{Diehl:2002nj}, but the
corresponding constraints depend on the values of all other couplings.
Moreover, the total event rate is likely dominated by systematic
errors, which we do not attempt to quantify here.

At present the physics case for transverse beam polarisation at a
linear collider is being discussed \cite{SLAC-proposal,POWER}.  Once
the planned degree of longitudinal polarisation is realised in
experiment, viz.\ about \mbox{$P_l^- = \pm 80\%$} for the electron and
about \mbox{$P_l^+ = \pm 60\%$} for the positron beam, the same
degrees of transverse polarisation $P_t^{\pm}$ are considered to be
feasible.  Then the important question arises of how to distribute the
available total luminosity among the different modes.  Thus, the
physics cases for the various polarisation modes must be studied.

The purpose of this paper is to analyse the gain or loss in
sensitivity to all 28 TGCs using transverse instead of longitudinal
beam polarisation in the reaction \mbox{$e^-e^+ \rightarrow W^-W^+
\rightarrow$ 4 fermions}.  To this end we consider the full normalised
event distribution.  Our work is complementary to
\cite{Fleischer:1993ix}, where the total cross section for different
$W$ boson helicities as well as the left-right and transverse
asymmetries---both integrated and as a function of the $W$ production
angle---were calculated for the same reaction, including order
$\alpha$ corrections and bremsstrahlung.  The sensitivity of the cross
section and of various angular distributions in the final state was
investigated in an early study of polarisation for LEP2
\cite{Zeppenfeld:1986na}.  Only a restricted number of $CP$ conserving
form factors without imaginary parts was considered in these works.
Here, in contrast, we determine the maximum sensitivity to the full
set of parameters by means of optimal observables.  The differential
cross section for arbitrary polarisation can be written as a sum where
the first term depends on $P_l^{\pm}$ and the second is proportional
to the product \mbox{$(P_t^- \cdot P_t^+)$}, see~(16) in
\cite{Diehl:2002nj}.  Hence, there can be only effects of transverse
polarisation if both beams are polarised and if both the electron and
the positron spin vectors have a transverse component.  In the
following we consider only longitudinal {\em or} transverse
polarisation, but no hybrid, though it is in general not excluded that
the sensitivity of the differential cross section to some parameters
can improve by considering more generic directions of the electron and
positron spin vectors.  Furthermore, we quantify the statement in
\cite{Diehl:2002nj} that the coupling \mbox{${\rm Im}(g_1^R +
\kappa_R)$} is measurable with transverse polarisation.

The outline of this work is as follows: In Sect.~\ref{sec-cross} we
summarise the results for the differential cross section with
arbitrary beam polarisation using the notation of \cite{Diehl:2002nj}.
We repeat the definitions of the tensors, frames and angles in detail
here, because they are crucial in the context of transverse
polarisation.  In Sect.~\ref{sec-initial} we recall the classification
of the TGCs into four symmetry classes.  We then show that this can be
exploited to measure couplings of different symmetry classes
independently with any type of beam polarisation (as it can be with
longitudinal polarisation, see \cite{Diehl:2002nj}).  In
Sect.~\ref{sec-sens} we make some general statements about the
measurement of TGCs in $W$ pair production with transverse beam
polarisation.  In Sect.~\ref{sec-resu} we present our numerical
results.  We give the minimum achievable errors on all 28 TGCs when
they are measured simultaneously, i.e.\ none of them is assumed to be
zero.  These results are then compared to the propects of measurement
with unpolarised beams and with longitudinal polarisation.  In
Sect.~\ref{sec-concl} we present our conclusions.


\section{\boldm{W} pair production}
\label{sec-cross}

In this section we introduce our notation and briefly review the
results of \cite{Diehl:2002nj} for the differential cross section of
the process
\begin{equation}
\label{eq:proc}
e^-e^+ \longrightarrow W^-W^+ \longrightarrow (f_1 \overline{f_2})
(f_3 \overline{f_4}),
\end{equation}
where the final state fermions are leptons or quarks.  All definitions
and results of this section can be found in \cite{Diehl:2002nj} and
are repeated here only for the convenience of the reader.  As in
\cite{Diehl:2002nj} the process is calculated in the SM with the
$\gamma WW$ and $ZWW$ vertices replaced by their most general form
allowed by Lorentz invariance.  In particular, we use the
fermion-boson vertices of the SM.  We parameterise the $\gamma WW$ and
$ZWW$ vertices by the parameters $g_1^V$, $\kappa_V$, $\lambda_V$,
$g_4^V$, $g_5^V$, $\tilde{\kappa}_V$, $\tilde{\lambda}_V$ with
\mbox{$V = \gamma$ or $Z$}, see (2.4) in \cite{Hagiwara:1986vm}.
In the SM at tree level one has
\begin{equation}
g_1^V = 1,\;\;\; \kappa_V = 1 \;\;\;\;\;\;\;(V = \gamma, Z),
\label{eq:coupsm}
\end{equation}
and all other couplings are equal to zero.  As usual we denote
deviations from the SM values (\ref{eq:coupsm}) by \mbox{$\Delta g_1^V
= g_1^V - 1$} and \mbox{$\Delta \kappa_V = \kappa_V - 1$}.  It has
been emphasised \cite{Diehl:1993br} that the following linear
combinations of couplings, introduced in
\cite{Hagiwara:1986vm}, can be measured with much smaller correlations:
\begin{eqnarray}
g_1^L & = & 4\sin^2\theta_W\, g_1^{\gamma} + (2 - 4\sin^2\!\theta_W )\,
\xi\, g_1^Z,    \nonumber \\
g_1^R & = & 4\sin^2\theta_W\, g_1^{\gamma} - 4\sin^2\!\theta_W\, \xi\,
g_1^Z,  
\label{eq:lrgz}
\end{eqnarray}
where $\xi = s/(s - m_Z^2)$, and similarly for the other couplings.
The L- and R-couplings respectively appear in the amplitudes for left-
and right-handed initial $e^-$.  We use the parameterisation
(\ref{eq:lrgz}) where appropriate.

\begin{figure}[h]
\center
\includegraphics[totalheight=5.1cm]{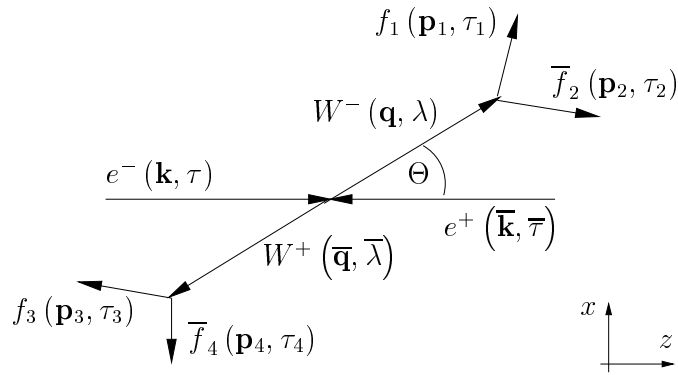}
\caption{\label{fig:pro}Momenta and helicities in the $e^-e^+$
c.m.~frame.}
\end{figure}

Fig.~\ref{fig:pro} shows our definitions of the particle momenta and
helicities.  The production of the $W$ bosons is described in the
$e^-e^+$ c.m.~frame.  The coordinate axes are chosen such that the
$e^-$~momentum points in the positive $z$-direction and the $y$~unit
vector is given by \mbox{$\boldm{\hat{e}_y} = ({\bf k} \times {\bf q})
/ |{\bf k} \times {\bf q}|$}.
At a given c.m.~energy $\sqrt{s}$, a pure initial state of $e^-$ and
$e^+$ with longitudinal polarisation can be uniquely specified by the
$e^-$ and $e^+$ helicities in the c.m.~frame:
\begin{equation}
|\tau \overline{\tau} \rangle = |e^-({\bf k},\tau) e^+(\overline{\bf
k}, \overline{\tau})
\rangle\;\;\;\;\;\; (\tau, \overline{\tau} = \pm 1).
\end{equation}
Here and in the following fermion helicity indices are normalised to
1.  A mixed initial state with general polarisation is given by the
operator
\begin{equation}
\boldm{\rho} = \sum_{(\tau)} |\tau \overline{\tau} \rangle
\rho_{(\tau \overline{\tau})(\tau' \overline{\tau}{}')} \langle \tau'
\overline{\tau}{}'|,
\label{eq:density}
\end{equation}
where $\rho_{(\tau \overline{\tau})(\tau' \overline{\tau}{}')}$ is the
spin density matrix of the combined $e^-e^+$ system and $(\tau)$
denotes summation over $\tau$, $\overline{\tau}$, $\tau'$ and
$\overline{\tau}{}'$.  The matrix entries satisfy
\begin{equation}
\rho^{\ast}_{(\tau \overline{\tau})(\tau' \overline{\tau}{}')}
= \rho^{\phantom{\ast}}_{(\tau' \overline{\tau}{}')(\tau
\overline{\tau})} , 
\qquad \qquad
\sum_{\tau, \overline{\tau}} \rho_{(\tau \overline{\tau})(\tau
\overline{\tau})} = 1 .
\end{equation}
The differential cross section with the initial state \boldm{\rho} is
given by the trace
\begin{equation}
\label{eq:trace}
{\rm d}\sigma |_{\rho} 
= {\textstyle \rm tr}(\boldm{{\rm d}\sigma \rho})
= \sum_{(\tau )} {\rm d}\sigma_{(\tau'
\overline{\tau}{}')(\tau \overline{\tau})}\, \rho_{(\tau
\overline{\tau})(\tau' \overline{\tau}{}')} ,
\end{equation}
where \boldm{{\rm d}\sigma} is the operator
\begin{equation}
\label{eq:dsdef}
\boldm{{\rm d}\sigma} = \sum_{(\tau )}
|\tau' \overline{\tau}{}' \rangle {\rm d}\sigma_{(\tau'
\overline{\tau}{}')(\tau \overline{\tau})} \langle \tau
\overline{\tau}|.
\end{equation}
Using the narrow-width approximation for the $W$s, the matrix entries
in (\ref{eq:dsdef}) are
\begin{eqnarray}
{\rm d}\sigma_{(\tau' \overline{\tau}{}')(\tau \overline{\tau})} & = & 
\frac{\beta}{(8\pi)^6 (m_W\Gamma_W )^2 \, s} \,
\sum_{(\lambda)} {\rm d}{\cal P}_{(\tau' \overline{\tau}{}')(\tau
\overline{\tau})}^{(\lambda \overline{\lambda})(\lambda'
\overline{\lambda}{}')} \, {\rm d}{\cal D}^{}_{\lambda' \lambda} \,
{\rm d}\overline{\cal D}_{\overline{\lambda}{}'
\overline{\lambda}}\;\,.
\end{eqnarray}
Here $m_W$ is the $W$~boson mass, $\Gamma_W$ its width, and $\beta =
(1 - 4 m_W^2 / s)^{1/2}$                                        its
velocity in the $e^-e^+$~c.m.~frame.  The sum $(\lambda)$ runs over
$\lambda$, $\lambda'$, $\overline{\lambda}$ and
$\overline{\lambda}{}'$.  Denoting the transition operator by ${\cal
T}$ the differential production tensor for the $W$~pair is
\begin{equation} 
{\rm d}{\cal P}_{(\tau' \overline{\tau}{}')(\tau 
\overline{\tau})}^{(\lambda 
\overline{\lambda})(\lambda' \overline{\lambda}{}')} =
{\rm d}(\cos\!\Theta)\:{\rm d}\Phi\;
\langle \lambda \overline{\lambda}, \Theta |
{\cal T} | \tau \overline{\tau} \rangle \, \langle \lambda'
\overline{\lambda}{}', \Theta | {\cal T} | \tau' \overline{\tau}{}'
\rangle^{\ast}
\label{eq:prod}
\end{equation}
and the tensors of the $W^-$ and $W^+$~decays in their respective
c.m.~frames are
\begin{eqnarray}
{\rm d}{\cal D}_{\lambda'\lambda} & = &
{\rm d}(\cos\!\vartheta)\,{\rm d}\varphi\;\langle f_1 \overline{f_2} |
{\cal T} | \lambda \rangle \, \langle f_1\overline{f_2} | {\cal T} |
\lambda' \rangle^{\ast},\nonumber \\
{\rm d}\overline{\cal D}_{\overline{\lambda}{}'\overline{\lambda}} & = &
{\rm d}(\cos\!\overline{\vartheta})\,{\rm d}\overline{\varphi}\;\langle f_3
\overline{f_4} | {\cal T} | \overline{\lambda} \rangle \, \langle
f_3\overline{f_4} | {\cal T} | \overline{\lambda}{}' \rangle ^{\ast}.
\label{eq:deca}
\end{eqnarray}
The matrix elements in (\ref{eq:prod}) and (\ref{eq:deca}) are given
in \cite{Hagiwara:1986vm}.  The $W$ bosons are produced by SM neutrino
exchange in the $t$-channel, and by $s$-channel photon and $Z$
production via the TGCs of the SM and the anomalous TGCs.  We note
that in our process there are possible effects of physics beyond the
SM which cannot be parameterised in terms of TGCs, see
e.g.~\cite{Alam:db}.  For further discussion of this point we refer to
\cite{Diehl:2002nj}.

\begin{figure}
\centering
\includegraphics[totalheight=4.3cm]{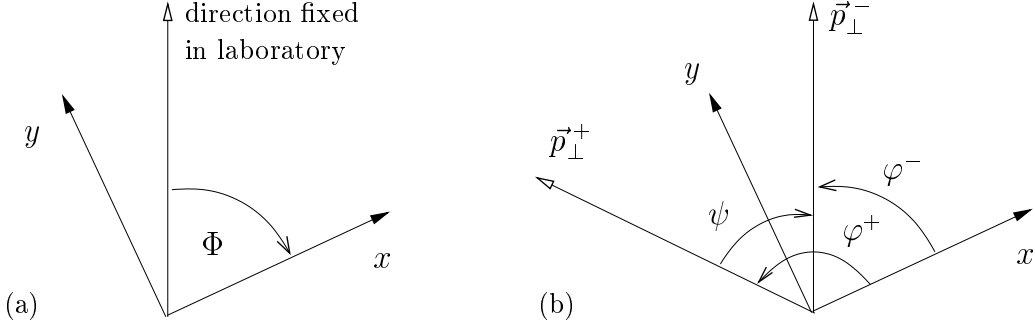}
\caption{\label{fig:angle}Definition of azimuthal angles.  The
$x$-axis points in the direction of $\bf{q}_{\,\perp}$.}
\end{figure}

The $W$ helicity states occurring on the right-hand side of
(\ref{eq:prod}) are defined in the frame of Fig.~\ref{fig:pro}.  By
$\Theta$ we denote the polar angle between the $W^-$ and $e^-$
momenta.  Furthermore, we choose a fixed direction transverse to the
beams in the laboratory.  By $\Phi$ we denote the azimuthal angle
between this fixed direction and the \mbox{$e^-e^+ \rightarrow
W^-W^+$}\ scattering plane (see Fig.~\ref{fig:angle}(a)).
The respective frames for the decay tensors (\ref{eq:deca}) are
defined by a rotation by $\Theta$ about the $y$-axis of the frame in
Fig.~\ref{fig:pro} (such that the $W^-$~momentum points in the new
positive $z$-direction) and a subsequent rotation-free boost into the
c.m.~system of the corresponding $W$.  The spherical coordinates in
(\ref{eq:deca}) are those of the $f_1$ and $\overline{f_4}$ momentum
directions, respectively.  In its rest frame, the quantum state of a
$W$~boson is determined by its polarisation.  In the narrow-width
approximation the intermediate $W$s are treated as on-shell, so that
we can take as basis the eigenstates of the spin operator $S_z$ with
the three eigenvalues \mbox{$\lambda = \pm 1,0$}.

Subsequently, we set the electron mass to zero.  Since we assume SM
couplings for the electron-boson vertices we have the relations
\begin{equation}
{\rm d}\sigma_{(\tau' \overline{\tau}{}')(\tau \overline{\tau})} = 0
     \qquad \qquad \mbox{for~} \tau = \overline{\tau} \mbox{~or~}
     \tau' = \overline{\tau}{}'.        
\label{eq:zero}
\end{equation}
Assuming the initial beams to be uncorrelated their spin density
matrix factorises:
\begin{equation}
\rho_{(\tau \overline{\tau})(\tau' \overline{\tau}{}')} = \rho_{\tau
\tau'} \overline{\rho}{}_{\, \overline{\tau}\,\overline{\tau}{}'}\;\,,
\label{eq:fact}
\end{equation}
where $\rho_{\tau \tau'}$ and
$\overline{\rho}{}_{\,\overline{\tau}\,\overline{\tau}{}'}$ are the
two Hermitian and normalised spin density matrices of $e^-$ and $e^+$,
respectively.  As in \cite{Diehl:2002nj} we parameterise these
matrices by
\begin{eqnarray}
\rho_{\tau \tau'} = \frac{1}{2} \left( \mathbbm1 + \vec{p}\,^- 
\! \cdot \vec{\sigma} \right)_{\tau \tau'}\,,\;\;\;\;\;
\overline{\rho}_{\,\overline{\tau}\,\overline{\tau}{}'} = \frac{1}{2}
\left( \mathbbm1 - \vec{p}\,^+ \! \cdot \vec{\sigma}^{\,\ast}
\right)_{\overline{\tau}\,\overline{\tau}{}'}
\label{eq:dens}
\end{eqnarray}
with
\begin{equation}
\vec{p}\,^{\pm} = P_t^{\pm} \left( \begin{array}{c} 
                        \cos{\varphi^{\pm}} \\
                        \sin{\varphi^{\pm}} \\
                        0
                        \end{array} \right) 
        + P_l^{\pm} \left( \begin{array}{c}
                        0 \\
                        0 \\
                        \mp 1
                        \end{array} \right).
\label{eq:polvec}
\end{equation}
The range of the azimuthal angles is $0 \le \varphi^{\pm} < 2 \pi$.
The components of $\vec{\sigma}$ are the Pauli matrices.  The degrees
$P_t^{\pm}$ of transverse and $P_l^{\pm}$ of longitudinal polarisation
obey the relations \mbox{$(P_t^{\pm})^2 + (P_l^{\pm})^2 \le 1$} and
\mbox{$P_t^{\pm} \ge 0$}.  The components of $\vec{p}\,^{\pm}$
in~(\ref{eq:polvec}) refer to the frame of \mbox{Fig.~\ref{fig:pro}}.
The relative azimuthal angle \mbox{$\psi = \varphi^- - \varphi^+$}
between $\vec{p}\,^-$ and $\vec{p}\,^+$ is fixed by the experimental
conditions, whereas the azimuthal angles $\varphi^-$ of $\vec{p}\,^-$
and $\varphi^+$ of $\vec{p}\,^+$ in the system of Fig.~\ref{fig:pro}
depend on the final state (see Fig.~\ref{fig:angle}(b)).  For the case
where \mbox{$P_t^- \ne 0$} we choose the transverse part of the vector
$\vec{p}\,^-$ as fixed direction in the laboratory.  Then we have
\mbox{$\Phi = - \varphi^-$}.  With these conventions the differential
cross section for arbitrary beam polarisation is\\
\parbox{\textwidth}{
\begin{eqnarray}
{\rm d}\sigma |_{\rho} & = & \frac{1}{4}\, \bigg\{ (1 + P_l^-)(1 - P_l^+)\,
 {\rm d}\sigma_{(+-)(+-)}
\label{eq:genpol} \\
 & & \hspace{0.3em} {}+ (1 - P_l^-)(1 + P_l^+)\, {\rm d}\sigma_{(-+)(-+)} 
\phantom{\bigg\{ \bigg\}}
\nonumber \\
 & & \hspace{0.3em} {}- 2 P_t^- P_t^+ 
        \Big[ \, {\rm Re}\, {\rm d}\sigma_{(+-)(-+)}\, \cos{(\psi + 2 \Phi)} 
\phantom{\bigg\{ \bigg\}}
\nonumber \\
 & & \hspace{4.2em}
             {}+ {\rm Im}\, {\rm d}\sigma_{(+-)(-+)}\, \sin{(\psi + 2 \Phi)}
 \Big] \bigg\}. 
\nonumber 
\end{eqnarray}}
In this formula the azimuthal angle $\Phi$ occurs only in the
arguments of the two trigonometric functions.


\section{\boldm{CP} and \boldm{CP\tilde{T}}}
\label{sec-initial}

As is well-known \cite{Hagiwara:1986vm} the real and the imaginary
parts of the couplings $g_1$, $\kappa$, $\lambda$, and $g_5$ are $CP$
conserving, whereas those of $g_4$, $\tilde{\kappa}$, and
$\tilde{\lambda}$ are $CP$ violating.  Here $CP$ denotes the combined
discrete symmetry of charge conjugation and parity transformation.
Furthermore, the real parts of the couplings are $CP\tilde{T}$
conserving, whereas their imaginary parts are $CP\tilde{T}$ violating,
where $\tilde{T}$ denotes ``na\"{\i}ve time reversal'', i.e.\ the
reversal of all particle momenta and spins without the interchange of
initial and final state.  Hence the TGCs can be classified as follows
\cite{Diehl:1993br,Diehl:1997ft}:\\\\
\begin{tabular}{rl}
~~~~(a) & $CP$ and $CP\tilde{T}$ conserving, \\
(b) & $CP$ conserving and $CP\tilde{T}$ violating, \\
(c) & $CP$ violating and $CP\tilde{T}$ conserving, \\
(d) & $CP$ and $CP\tilde{T}$ violating.
\end{tabular}\\\\
Since the interactions are invariant under rotations we can, instead
of a pure $CP\tilde{T}$ transformation, equally well consider
$R\,CP\tilde{T}$, i.e.\ $CP\tilde{T}$ followed by a rotation $R$ by
180 degrees around $\boldm{\hat{e}_y}$.  Notice that this differs from
the definition of $R$ in Sect.~3.3 of \cite{Diehl:2002nj} where the
rotation axis is \mbox{$({\bf k} \times \vec{p}\,^-)$}.  More detail
is given in Appendices \ref{app-phase} and~\ref{app-disc}.

When measured with an appropriate set of observables
\cite{Diehl:1993br}, couplings from different symmetry groups have
uncorrelated statistical errors (to leading order in the anomalous
couplings), provided that phase space cuts, detector acceptance and
the initial state are invariant under $CP$ and $R\,CP\tilde{T}$.  As
mentioned in \cite{Diehl:2002nj}, $CP$ and $R\,CP\tilde{T}$ violating
effects from the initial state in $e^-e^+ \rightarrow W^-W^+$ are
suppressed by $(m_e/m_W)$ with transverse beam polarisation and by
$(m_e/m_W)^2$ with longitudinal polarisation or unpolarised beams.
Consequently, these effects vanish in the limit $m_e \rightarrow 0$
for an arbitrary spin density matrix \boldm{\rho}.  This means that,
although the initial state is \emph{not} invariant under $CP$ and
$R\,CP\tilde{T}$, it is {\em effectively} invariant for our reaction
in the $m_e \rightarrow 0$ limit.  Let us make this more explicit.

Both the $CP$ and the $R\,CP\tilde{T}$ transformation of the initial
state leave the particle momenta unchanged and correspond to the
following transformation of the spin density matrix:
\begin{equation}
\rho_{(\tau \overline{\tau})(\tau' \overline{\tau}')}
\;\;\; \begin{CD}@>{CP,\;R\,CP\tilde{T}}>>\end{CD} \;\;\; 
\rho_{(\overline{\sigma} \sigma)(\overline{\sigma}' \sigma')} \; 
\epsilon_{\overline{\sigma}\,\overline{\tau}} \; 
\epsilon_{\sigma \tau} \; \epsilon_{\overline{\sigma}' \overline{\tau}'} 
\; \epsilon_{\sigma' \tau'}\;,
\label{eq:trans}
\end{equation}
where
\begin{equation}
\epsilon = \left( \begin{array}{rr} 0 & \phantom{-}1 \\ -1 & 0 \end{array} \right).
\label{eq:epsdef}
\end{equation}
This transformation rule is derived in Appendix~\ref{app-disc}.  Thus
invariance of the spin density matrix under either of the two
symmetries requires
\begin{equation}
\rho_{(\tau \overline{\tau})(\tau' \overline{\tau}')} =
\rho_{(\overline{\sigma} \sigma)(\overline{\sigma}' \sigma')} \; 
\epsilon_{\overline{\sigma}\,\overline{\tau}} \;
\epsilon_{\sigma \tau} \; \epsilon_{\overline{\sigma}' \overline{\tau}'} 
\; \epsilon_{\sigma' \tau'}\;.
\label{eq:invariance}
\end{equation}
If the spin density matrix factorises as in (\ref{eq:fact}) we find
\begin{equation}
\rho_{\tau \tau'} = \left( \epsilon^{\rm T} \overline{\rho} \, 
\epsilon \right)_{\tau \tau'}.
\label{eq:rhofact}
\end{equation}
In our parameterisation the spin density matrices are explicitly given
by
{
\renewcommand{\arraystretch}{1.3}
\begin{eqnarray}
\rho_{\tau \tau'} & = & \frac{1}{2}\left(\begin{array}{cc}
(1+P_l^-) & P_t^- e^{-{\rm i}\varphi^-} \\
P_t^- e^{{\rm i}\varphi^-} & (1-P_l^-)
\end{array}\right)_{\tau \tau'},\\[.5cm]
\label{eq-defrbar}
\overline{\rho}{}_{\,\overline{\tau}\,\overline{\tau}'} & = &
\frac{1}{2}\left(\begin{array}{cc}
(1+P_l^+) & -P_t^+ e^{{\rm i}\varphi^+} \\
-P_t^+ e^{-{\rm i}\varphi^+} & (1-P_l^+)
\end{array}\right)_{\overline{\tau}\,\overline{\tau}'}.
\end{eqnarray}
The requirement (\ref{eq:rhofact}) thus reads   
\begin{eqnarray}
\rho_{\tau \tau'} & = & \frac{1}{2}\left(\begin{array}{cc}
(1-P_l^+) & P_t^+ e^{-{\rm i}\varphi^+} \\
P_t^+ e^{{\rm i}\varphi^+} & (1+P_l^+)
\end{array}\right)_{\tau \tau'} ,
\end{eqnarray}
which leads to the following conditions on the polarisation parameters:
}   
\begin{equation}
P_l^- = - P_l^+,\;\;\;\;\;\;\;\; P_t^- = P_t^+,\;\;\;\;\;\;\;\;
\varphi^- = \varphi^+.
\label{eq:parcond}
\end{equation}
If we do not demand $CP$ or $R\,CP\tilde{T}$ invariance of the full
spin density matrix but only consider those matrix entries that give
non-vanishing amplitudes we find, instead of (\ref{eq:invariance}):
\begin{equation}
\tilde{\rho}_{(\tau \overline{\tau})(\tau' \overline{\tau}')} =
\tilde{\rho}_{(\overline{\sigma} \sigma)(\overline{\sigma}' \sigma')}
\; \epsilon_{\overline{\sigma}\,\overline{\tau}} \;
\epsilon_{\sigma \tau} \; \epsilon_{\overline{\sigma}' \overline{\tau}'} 
\; \epsilon_{\sigma' \tau'}\;,
\label{eq:cpinvred}
\end{equation}
with a ``reduced'' spin density matrix
\begin{equation}
\tilde{\rho}_{(\tau \overline{\tau})(\tau' \overline{\tau}')} = \left\{
\begin{array}{cl} \rho_{(\tau \overline{\tau})(\tau'
\overline{\tau}')} & {\rm for}\;\; \overline{\tau} = - \tau \;\; {\rm
and} \;\; \overline{\tau}' = - \tau'\\[.3cm]
0 & {\rm else.} \end{array} \right.
\label{eq:redspindens}
\end{equation}
Inserting this definition into (\ref{eq:cpinvred}) we find that the
condition for $CP$ or $R\,CP\tilde{T}$ invariance is trivially
fulfilled.  Under the assumption that only amplitudes with
\mbox{$\overline{\tau} = - \tau$} contribute to the differential cross
section and that the factorisation (\ref{eq:fact}) is possible, no
condition has therefore to be imposed on $P_l^{\pm}$, $P_t^{\pm}$ or
$\varphi^{\pm}$ to guarantee absence of $CP$ or $R\,CP\tilde{T}$
violation in the initial state.  Any violation of $CP$ or
$CP\tilde{T}$ is then due to the interaction.


\section{Sensitivity to couplings}
\label{sec-sens}

To investigate the prospects of measuring anomalous TGCs in the
process (\ref{eq:proc}) with transverse beam polarisation we use
optimal observables \cite{Atwood:1991ka,Diehl:1993br}.  In the limit
of small couplings, the statistical errors on the couplings determined
with this method are minimal compared with any other set of
observables.  To be more precise, these observables minimise the
errors for a given {\em normalised} event distribution.  The
integrated cross section still contains complementary information on
the couplings.  For more details concerning optimal observables we
refer to Sect.~3 in \cite{Diehl:2002nj} where they are applied to the
same reaction with longitudinal beam polarisation.  Before listing our
results in Sect.~\ref{sec-resu} we now discuss some aspects of
transverse polarisation in the context of the process (\ref{eq:proc}).

Looking at the differential cross section (\ref{eq:genpol}) we see
that a change in $\psi$ by \mbox{$\Delta \psi$} is equivalent to a
rotation of the whole event distribution about the beam axis by
\mbox{$\Delta \Phi = \Delta \psi/2$}.  It neither changes the shape of
the distribution nor the total event rate.  The sensitivity to the
TGCs thus does not depend on $\psi$.  Integrating the differential
cross section over $\Phi$, the terms proportional to \mbox{$\cos{(\psi
+ 2 \Phi)}$} and \mbox{$\sin{(\psi + 2 \Phi)}$} in (\ref{eq:genpol})
vanish.  The total cross section is hence independent of $P_t^-$,
$P_t^+$ and $\psi$.  Therefore, in absence of longitudinal
polarisation, the total cross section with transversely polarised
beams equals that with unpolarised beams.  This cross section is shown
in Fig.~4 of \cite{Diehl:2002nj} for the SM and with various anomalous
TGCs.  Some other quantities required for the optimal observable
method are also the same for pure transverse polarisation and for
unpolarised beams.  These are in particular the total cross section in
the SM $\sigma_0$, the expectation values of the optimal observables
in the SM $E_0[{\cal O}_i]$, and the normalised second-order part of
the total cross section $\hat{\sigma}_{2ij}$, see~(26) and (36) in
\cite{Diehl:2002nj}.

As seen in Sect.~\ref{sec-initial}, the initial state is not invariant
under the discrete symmetries $CP$ and $R\,CP\tilde{T}$ for generic
beam polarisation.  It is however {\em effectively} invariant if the
electron mass is neglected, because then only a subset of helicity
amplitudes is non-zero.  Hence the results of Sect.~3.3 in
\cite{Diehl:2002nj} apply, i.e.\ a given optimal observable is
sensitive only to couplings of the same symmetry class (a), (b), (c),
or (d).  Measurement errors on couplings of different symmetry classes
are not correlated to leading order in the anomalous couplings.
Furthermore, the first-order terms in the integrated cross section
vanish except for symmetry (a), where only the $g_5^R$-term is zero.


\section{Numerical results}
\label{sec-resu}

In this section we present our results for the sensitivity to
anomalous TGCs in the reaction (\ref{eq:proc}) with transverse beam
polarisations \mbox{$P_t^- = 80\%$} of the electron and \mbox{$P_t^+ =
60\%$} of the positron beam.  As in \cite{Diehl:2002nj} we consider
only events where one $W$ boson decays into a quark-antiquark pair and
the other one into $e\nu$ and $\mu \nu$.  These decay channels have a
branching ratio of altogether $8/27$.  We assume that the two jets of
the hadronic $W$ decay cannot be identified as originating from the
up- and down-type (anti)quark.  This must be taken into account in the
definition of the optimal observables as explained in
\cite{Diehl:1997ft}.  To measure the coupling parameters these
observables have the maximum sensitivity that one can obtain from the
sum of the event distributions corresponding to the two final states.

For the masses of the $W$ and $Z$ we use \mbox{$m_W = 80.42$~GeV} and
\mbox{$m_Z = 91.19$~GeV} \cite{Hagiwara:fs} and define the weak mixing
angle by $\sin^2\!\theta_W = 1 - m_W^2/m_Z^2$.  For the total event
rate $N$ with transverse beam polarisation we use the values listed in
Table~3 of \cite{Diehl:2002nj}, viz.\ \mbox{$1.14 \times 10^6$} for a
c.m.~energy of 500~GeV and \mbox{$1.19 \times 10^6$} for 800~GeV.
These values are computed for an effective electromagnetic coupling
$\alpha = 1/128$ and integrated luminosities of 500~${\rm fb}^{-1}$ at
500~GeV and 1~${\rm ab}^{-1}$ at 800~GeV.

In Tables~\ref{tab:sensa} to \ref{tab:sensd} we give the standard
deviations \mbox{$\delta h_i = [V(h)_{ii}]^{1/2}$} for the couplings
of symmetry classes (a) to (d), as well as the correlation matrices
\begin{equation}
W(h)_{ij} = \frac{V(h)_{ij}}{\sqrt{V(h)_{ii} V(h)_{jj}}},
\label{eq:wdef}
\end{equation}
where $V(h)_{ij}$ is the covariance matrix of the couplings in the
L-R-parameterisation~(\ref{eq:lrgz}).  $V$ and $W$ are evaluated with
zero anomalous couplings, and errors on couplings in different
symmetry classes are uncorrelated to this accuracy.  The $\delta h_i$
are the errors obtained without assuming any other anomalous coupling
to be zero.  For symmetry~(b) we use the linear combinations
\mbox{$\tilde{h}_{\pm} = {\rm Im}(g_1^R \pm \kappa_R)/\sqrt{2}$}
instead of ${\rm Im}\,g_1^R$ and ${\rm Im}\,\kappa_R$ to allow for
better comparison with the results for unpolarised beams and
longitudinal polarisation, where the normalised event distribution is
sensitive to $\tilde{h}_-$, but not to $\tilde{h}_+$.  The range of
the $\delta h_i$ within each symmetry class is from about \mbox{$5
\times 10^{-4}$} to about \mbox{$5 \times 10^{-3}$}.  Notice that both
$\tilde{h}_+$ and $\tilde{h}_-$ are measurable with an error of about
\mbox{$3.5 \times 10^{-3}$} using transverse polarisation.  This
confirms and makes quantitive the result of \cite{Diehl:2002nj} that
one is indeed sensitive to $\tilde{h}_+$ with transverse polarisation.
Also the sensitivity to $\tilde{h}_-$ is significantly better than
with unpolarised beams, where the error is about $10^{-2}$.  The high
correlation between $\tilde{h}_+$ and $\tilde{h}_-$ however suggests
that the parameterisation with \mbox{${\rm Im}\,g_1^R$} and
\mbox{${\rm Im}\,\kappa_R$} is preferable in an analysis of the data
from transverse polarisation (whereas it is inadequate with
longitudinal polarisation or unpolarised beams).  The gain by
different types of polarisation at 500~GeV can be seen from
Tables~\ref{tab:pol1} to~\ref{tab:pol2im} for the four symmetry
classes.  In Tables~\ref{tab:pol3} to~\ref{tab:pol4im} the same is
shown for 800~GeV.  To allow for better comparison with other studies
we use the photon- and $Z$-couplings for the results of symmetries
(a), (c) and (d) instead of the L- and R-couplings, although the
latter are in general less correlated.   We use however the
L-R-couplings for symmetry (b), where only one coupling is
unmeasurable without transverse beam polarisation.  In the
$\gamma$-$Z$-parameterisation, four couplings,
\mbox{Im$\,g_1^{\gamma}$}, \mbox{Im$\,g_1^Z$},
\mbox{Im$\,\kappa_{\gamma}$} and \mbox{Im$\,\kappa_Z$}, are not
measurable in the absence of transverse polarisation, because their
linear combination $\tilde{h}_+$ is not.  In the unpolarised case the
assumed luminosity is 500~fb$^{-1}$ at 500~GeV and 1~ab$^{-1}$ at
800~GeV.  The same values are used for the results with transverse
polarisation in the fourth row of each table.  For the results with
longitudinal $e^-$ polarisation in the second row we assume that one
half of the luminosity is used for \mbox{$P_l^- = +80\%$} and the
other half for \mbox{$P_l^- = -80\%$}.  Similarly, for the results in
the third row with additional longitudinal $e^+$ polarisation we
assume that the total luminosity is equally distributed among the
settings with \mbox{$(P_l^-,P_l^+) = (+80\%,-60\%)$} and
$(-80\%,+60\%)$.  For each of rows number two and three, the results
from the two settings are combined in the conventional way, i.e.\ we
take the two covariance matrices $V_1$ and $V_2$, and compute the
matrix
\begin{equation}
\label{eq-vcomb}
V = \left(V_1^{-1} + V_2^{-1}\right)^{-1}.
\end{equation}
This is the covariance matrix on the couplings if they are determined
by a weighted average from two individual measurements.  $V_1$, $V_2$
and $V$ are 8$\times$8 matrices for symmetry class (a) and 6$\times$6
matrices for symmetry classes (c) and (d), whereas in the case of
symmetry class (b) they are 7$\times$7 matrices since the coupling
$\tilde{h}_+$ is excluded.  The square roots of the diagonal elements
of $V$ are then the 1$\sigma$-errors, which we list in the second and
third rows of Tables~\ref{tab:pol1} to \ref{tab:pol4im}.

For a c.m.~energy of 500~GeV the errors with unpolarised beams are
between $10^{-3}$ and $10^{-2}$ in the $\gamma$-$Z$-parameterisation
(see Tables~\ref{tab:pol1}, \ref{tab:pol2}, and~\ref{tab:pol2im}).
All errors (with or without polarisation) are smaller at 800~GeV (see
Tables~\ref{tab:pol3} to~\ref{tab:pol4im}), notably for \mbox{${\rm
Re}\, \Delta \kappa_{\gamma}$} and \mbox{${\rm Im}\,\lambda_R$}.  For
both c.m.~energies the errors on all couplings in the
$\gamma$-$Z$-parameterisation are about a factor 2 smaller with
longitudinal $e^-$ polarisation and unpolarised $e^+$~beam compared to
the case where both beams are unpolarised.  With additional
longitudinal $e^+$ polarisation this factor is between 3 and 4 for all
couplings, except for ${\rm Re}\, \Delta \kappa_Z$ at 800~GeV where it
is 4.7.  If both beams have transverse polarisation, the errors on
most couplings are approximately of the same size as in the situation
where only the $e^-$ beam has longitudinal polarisation.  Only for
${\rm Re}\,\lambda_{\gamma}$, ${\rm Re}\,\lambda_Z$, ${\rm
Re}\,\tilde{\lambda}_{\gamma}$ and ${\rm Re}\,\tilde{\lambda}_Z$ are
they smaller, viz.\ they are of the same size as with both beams
longitudinally polarised.  This is true for both energies.  If
electron as well as positron polarisation is available we thus
conclude that, regarding the 1$\sigma$-standard deviations on the TGCs
(without assuming any coupling to be zero) {\em longitudinal}
polarisation is the preferable choice, except for $\tilde{h}_+$.  We
emphasize that we are better with longitudinal polarisation also for
the $CP$ violating couplings ${\rm Re}\,g_4^V$, ${\rm
Re}\,\tilde{\lambda}_V$ and ${\rm Re}\,\tilde{\kappa}_V$ with \mbox{$V
= \gamma$ or $Z$}.

Furthermore, we analyse how correlations between couplings depend on
beam polarisation.  Given the large number of parameters, small
correlations are highly desirable.  For brevity we do not present the
full correlation matrices here for all different types of polarisation
but only give the average over the absolute values of the off-diagonal
elements in the correlation matrices (see Table \ref{tab:corr}).
Furthermore, we restrict ourselves to symmetry (a) and a c.m.~energy
of 500~GeV.

Apart from the average over all 28 matrix entries we list the averages
over the correlations between L-couplings, between R-couplings and
those between one L- and one R-coupling.  We see that no type of
polarisation changes the average correlation between two L-couplings
significantly.  The average correlation between the R-couplings is
most advantageous for transverse polarisation (26\%), whereas in the
other cases it ranges from $37\%$ to $42\%$.  On the other hand the
L-couplings are hardly correlated with the R-couplings for
longitudinal polarisation of $e^-$ and $e^+$ (2\%).  This deteriorates
with transverse polarisation, but the correlations remain very
small~(8\%).  Altogether, regarding the size of the correlations there
is no strong argument to prefer one type of polarisation or the other.

Finally, we remark that the sensitivity to TGCs in our reaction has
been analysed in \cite{Menges:2001gg} for unpolarised beams and for
longitudinal polarisation.  A maximum number of five $CP$ conserving
and four $CP$ violating couplings was considered, but no imaginary
parts were included (see Tables~5 and~6 of \cite{Menges:2001gg}). The
author used a spin density matrix method where statistical errors are
not necessarily optimal.  A direct comparison with our results is
however not possible. On the one hand the multi-parameter analysis of
\cite{Menges:2001gg} includes beamstrahlung, initial state radiation
and non-resonant diagrams.  For the single parameter fits the full
background calculated with PYTHIA and also detector acceptance is
included.  On the other hand only a restricted number of couplings is
considered.  An analysis using optimal observables with a full
detector simulation and all 28 couplings would be desirable for
unpolarised beams and both types of polarisation.  This is however
beyond the scope of our present work.


\section{Conclusions}
\label{sec-concl}

\suppressfloats

We have studied the prospects of measuring TGCs at a future linear
collider in $W$ pair production with transverse beam polarisation.
Effects due to transverse polarisation can only occur if both beams
are polarised and if both spins have a transverse component.  The
classification of the TGCs into four groups according to their
properties under the discrete symmetries $CP$ and $CP\tilde{T}$
remains valid in the case of transverse polarisation, neglecting
effects which are at most \mbox{$O(m_e/m_W)$}.  Using optimal
observables these four groups of parameters can be measured without
statistical correlation.  Within each group, the errors on TGCs are
correlated.  We have given the errors and correlation matrices for a
c.m.~energy of 500~GeV with transverse polarisation of the electron
\mbox{($P_t^- = 80\%$)} and the positron beam \mbox{($P_t^+ = 60\%$)}
in the parameterisation with L- and R-couplings.  The errors range
from about \mbox{$5 \times 10^{-4}$} to about \mbox{$5 \times
10^{-3}$}.  Moreover, we have compared the errors---mainly in the
$\gamma$-$Z$-parameterisation---for transverse polarisation with those
for unpolarised beams and with those for one or both beams
longitudinally polarised.  For most couplings the errors obtainable
with transverse polarisation are of the same order as with
longitudinal $e^-$ polarisation and unpolarised $e^+$ beam.  If one
has both beams polarised and can choose between longitudinal or
transverse polarisation it is therefore advantageous to use
longitudinal polarisation, except for the measurement of
$\tilde{h}_+$.  For the real parts of the couplings, the only
advantage of transverse polarisation we found is that the average
correlation among R-couplings is slightly reduced.  The coupling
$\tilde{h}_+$, however, is unmeasurable from the normalised event
distribution with longitudinal polarisation, but it can be measured
with an error of \mbox{$3.2 \times 10^{-3}$} using transverse
polarisation.  This suggests to use some fraction of the total
luminosity to run the collider in this mode in order to get a
comprehensive measurement of all TGCs.  The required luminosity for a
certain desired value of the statistical error on \mbox{${\rm
Im}(g_1^R + \kappa_R)/\sqrt{2}$} can be easily calculated by applying
the corresponding statistical factor.


\section*{Acknowledgements}

The authors are grateful to P.~Bock and G.~Moortgat-Pick for useful
remarks.  This work was supported by the German Bundesministerium
f\"ur Bildung und Forschung, project~no.~05HT9HVA3, and the Deutsche
Forschungsgemeinschaft with the Graduiertenkolleg ``Physikalische
Systeme mit vielen Freiheitsgraden'' in Heidelberg.


\begin{table}
\caption{\label{tab:sensa} Errors $\delta h$ in units of $10^{-3}$ on
the couplings of symmetry (a) (see Sect.~\ref{sec-initial}) in the
presence of all anomalous couplings, and correlation matrix $W(h)$ at
$\sqrt{s}=$~500~GeV with transverse beam polarisation
\mbox{$(P_t^-,P_t^+) = (80\%,60\%)$}.}
\begin{center}
\leavevmode
\footnotesize
\begin{tabular}{l|r|rrrr|rrrr}
$\;\;\;h$ & $\delta h \times 10^3$ & Re$\,\Delta g_1^L$ & Re$\,\Delta
\kappa_L$ & Re$\,\lambda_L$ & Re$\,g_5^L$ & Re$\,\Delta g_1^R$ & Re$\,\Delta
\kappa_R$ & Re$\,\lambda_R$ & Re$\,g_5^R$ \\
\hline
&&&&&&&&&\\[-2.5ex]
Re$\,\Delta g_1^L$ & $2.5$ & 1 & $-0.58$ & $-0.36$ & $0.17$ & $-0.068$ &
$0.18$ & $-0.011$ & $0.11$\\
Re$\,\Delta \kappa_L$ & $0.72$ && 1 & $0.077$ & $0.013$ & $0.075$ &
$-0.46$ & $0.023$ & $-0.014$\\
Re$\,\lambda_L$ & $0.58$ &&& 1 & $-0.011$ & $0.053$ & $-0.0040$ &
$0.029$ & $0.045$\\
Re$\,g_5^L$ & $2.0$ &&&& 1 & $-0.14$ & $-0.0027$ & $-0.038$ & $0.085$\\
\hline
&&&&&&&&&\\[-2.5ex]
Re$\,\Delta g_1^R$ & $4.2$ &&&&& 1 & $-0.56$ & $-0.41$ & $0.35$\\
Re$\,\Delta \kappa_R$ & $1.2$ &&&&&& 1 & $0.075$ & $-0.086$\\
Re$\,\lambda_R$ & $0.99$ &&&&&&& 1 & $-0.066$\\
Re$\,g_5^R$ & $3.5$ &&&&&&&& 1
\end{tabular}
\end{center}
\end{table}


\begin{table}
\caption{\label{tab:sensb} Same as Table~\ref{tab:sensa}, but for
symmetry~(b).  We use the abbreviations \mbox{$\tilde{h}_{\pm} = {\rm
Im}(g_1^R \pm \kappa_R)/\sqrt{2}$}.}
\begin{center}
\leavevmode
\footnotesize
\begin{tabular}{l|r|rrrr|rrrr}
$\;\;\;h$ & $\delta h \times 10^3$ & Im$\,g_1^L$ & Im$\,\kappa_L$ &
Im$\,\lambda_L$ & Im$\,g_5^L$ & $\tilde{h}_-$ & $\tilde{h}_+$ & 
Im$\,\lambda_R$ &
Im$\,g_5^R$\\
\hline
&&&&&&&&&\\[-2.5ex]
Im$\,g_1^L$ & $2.6$ & 1 & $-0.63$       & $-0.49$ & $-0.20$ & $0.050$ &
$-0.037$ & $0.061$ & $0.028$\\
Im$\,\kappa_L$ & $1.2$ && 1 & $0.19$ & $0.14$ & $-0.072$ & $0.051$ &
$-0.029$ & $0.22$\\
Im$\,\lambda_L$ & $0.46$ &&& 1 & $0.015$ & $0.024$ & $0.048$ &
$-0.063$ & 
$-0.089$\\
Im$\,g_5^L$ & $2.0$ &&&& 1 & $-0.063$ & $-0.053$ & $0.10$ & $0.18$\\
\hline
&&&&&&&&&\\[-2.5ex]
$\tilde{h}_-$ & $3.7$ &&&&& 1 & $0.81$ & $-0.39$ & $0.16$\\
$\tilde{h}_+$ & $3.2$ &&&&&& 1 & $-0.39$ & $0.11$\\
Im$\,\lambda_R$ & $0.98$ &&&&&&& 1 & $-0.0041$\\
Im$\,g_5^R$ & $4.4$ &&&&&&&& 1 
\end{tabular}\\
\end{center}
\end{table}


\begin{table}
\caption{\label{tab:sensc} Same as Table~\ref{tab:sensa}, but for
symmetry~(c).}
\begin{center}
\leavevmode
\footnotesize
\begin{tabular}{l|r|rrr|rrr}
$\;\;\;h$ & $\delta h \times 10^3$ & Re$\,g_4^L$ & Re$\,\tilde{\lambda}_L$
& Re$\,\tilde{\kappa}_L$ & Re$\,g_4^R$ & Re$\,\tilde{\lambda}_R$ &
Re$\,\tilde{\kappa}_R$\\
\hline
&&&&&&&\\[-2.5ex]
Re$\,g_4^L$ & $2.4$ & 1 & $-0.0082$ & $-0.50$ & $-0.072$ & $-0.079$ & 
$0.084$\\
Re$\,\tilde{\lambda}_L$ & $0.58$ && 1 & $0.30$ & $0.022$ & $0.030$ & 
$-0.074$\\
Re$\,\tilde{\kappa}_L$ & $2.6$ &&& 1 & $0.090$ & $0.056$ & $0.063$\\
\hline
&&&&&&&\\[-2.5ex]
Re$\,g_4^R$ & $3.9$ &&&& 1 & $-0.013$ & $-0.11$\\
Re$\,\tilde{\lambda}_R$ & $0.99$ &&&&& 1 & $0.41$\\
Re$\,\tilde{\kappa}_R$ & $4.1$ &&&&&& 1 
\end{tabular}
\end{center}
\end{table}


\begin{table}
\caption{\label{tab:sensd} Same as Table~\ref{tab:sensa}, but for
symmetry~(d).}
\begin{center}
\leavevmode
\footnotesize
\begin{tabular}{l|r|rrr|rrr}
$\;\;\;h$ & $\delta h \times 10^3$ & Im$\,g_4^L$ & Im$\,\tilde{\lambda}_L$
& Im$\,\tilde{\kappa}_L$ & Im$\,g_4^R$ & Im$\,\tilde{\lambda}_R$ &
Im$\,\tilde{\kappa}_R$\\
\hline
&&&&&&&\\[-2.5ex]
Im$\,g_4^L$ & $1.8$ & 1 & $0.0044$ & $0.19$ & $0.11$ & $0.086$ & 
$-0.0072$\\
Im$\,\tilde{\lambda}_L$ & $0.45$ && 1 & $0.51$ & $-0.10$ & $-0.056$ & 
$-0.022$\\
Im$\,\tilde{\kappa}_L$ & $1.9$ &&& 1 & $-0.18$ & $-0.047$ & $0.0037$\\
\hline
&&&&&&&\\[-2.5ex]
Im$\,g_4^R$ & $3.6$ &&&& 1 & $-0.021$ & $-0.32$\\
Im$\,\tilde{\lambda}_R$ & $0.97$ &&&&& 1 & $0.43$\\
Im$\,\tilde{\kappa}_R$ & $3.7$ &&&&&& 1 
\end{tabular}
\end{center}
\end{table}


\begin{table}
\caption{\label{tab:pol1} Errors $\delta h$ in units of $10^{-3}$ on
the couplings of symmetry (a) in the presence of all anomalous
couplings at \mbox{$\sqrt{s}=$ 500 GeV}, with unpolarised beams and
with different beam polarisations.}
\begin{center}
\leavevmode
\footnotesize
\begin{tabular} {l|rrrrrrrr}
 & Re$\,\Delta g_1^{\gamma}$ & Re$\,\Delta g_1^Z$ & Re$\,\Delta
\kappa_{\gamma}$ & Re$\,\Delta \kappa_Z$ & Re$\,\lambda_{\gamma}$ &
Re$\,\lambda_Z$ & Re$\,g_5^{\gamma}$ & Re$\,g_5^Z$\\
\hline 
&&&&&&&&\\[-2.5ex]
 no polarisation & 6.5 & 5.2 & 1.3 & 1.4 & 2.3 & 1.8 & 4.4 & 3.3\\
$(P_l^-,P_l^+)=(\mp 80\%,0)$ & 3.2 & 2.6 & 0.61 & 0.58 & 1.1 & 0.86 & 
2.2 & 1.7\\
$(P_l^-,P_l^+)=(\mp 80\%,\pm 60\%)$ & 1.9 & 1.6 & 0.40 & 0.36 & 0.62 & 
0.50 & 1.4 & 1.1\\
$(P_t^-,P_t^+)=(80\%,60\%)$ & 2.8 & 2.4 & 0.69 & 0.82 & 0.69 & 0.55 & 
2.5 & 1.9
\end{tabular}
\normalsize
\end{center}
\end{table}


\begin{table}
\caption{\label{tab:pol1imLR} Same as Table~\protect\ref{tab:pol1},
but for symmetry~(b) and with the L-R-parameterisation.  We write
again \mbox{$\tilde{h}_{\pm} = {\rm Im}(g_1^R \pm
\kappa_R)/\sqrt{2}$}.  Using this parameterisation, a maximum number
of couplings can be measured without transverse beam polarisation.  In
the $\gamma$-$Z$-parameterisation, the four couplings
Im$\,g_1^{\gamma}$, Im$\,g_1^Z$, Im$\,\kappa_{\gamma}$ and
Im$\,\kappa_Z$ are not measurable without transverse polarisation.}
\begin{center}
\leavevmode
\footnotesize
\begin{tabular} {l|rrrr|rrrr}
 & Im$\,g_1^L$ & Im$\,\kappa_L$ & Im$\,\lambda_L$ & Im$\,g_5^L$ &
$\tilde{h}_-$ & $\tilde{h}_+$ & Im$\,\lambda_R$ & Im$\,g_5^R$\\
\hline   
&&&&&&&&\\[-2.5ex] 
no polarisation & 2.7 & 1.7 & 0.48 & 2.5 & 11 & --- & 3.1 & 17 \\ 
$(P_l^-,P_l^+)=(\mp 80\%,0)$ & 2.6 & 1.2 & 0.45 & 2.0 & 4.5 & --- &
1.4 & 
4.3 \\ 
$(P_l^-,P_l^+)=(\mp 80\%,\pm 60\%)$ & 2.1 & 0.95 & 0.37 & 1.6 & 2.5 &
--- & 0.75 & 2.3 \\ 
$(P_t^-,P_t^+)=(80\%,60\%)$ & 2.6 & 1.2 & 0.46 & 2.0 & 3.7 & 3.2 &
0.98 & 4.4
\end{tabular}
\normalsize
\end{center}
\end{table}


\begin{table}
\caption{\label{tab:pol2} Same as Table \protect\ref{tab:pol1}, but
for symmetry~(c).}
\begin{center}
\leavevmode
\footnotesize
\begin{tabular} {l|rrrrrr}
 & Re$\,g_4^{\gamma}$ & Re$\,g_4^Z$ & Re$\,\tilde{\lambda}_{\gamma}$ &
Re$\,\tilde{\lambda}_Z$ & Re$\,\tilde{\kappa}_{\gamma}$ & Re$\,\tilde{\kappa}_Z$\\
\hline
&&&&&&\\[-2.5ex]
 no polarisation & 6.2 & 5.1 & 2.4 & 1.9 & 7.3 & 5.4\\
$(P_l^-,P_l^+)=(\mp 80\%,0)$ & 3.0 & 2.5 & 1.1 & 0.90 & 3.4 & 2.7\\
$(P_l^-,P_l^+)=(\mp 80\%,\pm 60\%)$ & 1.8 & 1.5 & 0.64 & 0.52 & 2.1 & 1.7\\
$(P_t^-,P_t^+)=(80\%,60\%)$ & 2.7 & 2.3 & 0.69 & 0.55 & 2.9 & 2.3
\end{tabular}
\normalsize
\end{center}
\end{table}


\begin{table}
\caption{\label{tab:pol2im} Same as Table \protect\ref{tab:pol1}, but
for symmetry~(d).}
\begin{center}
\leavevmode
\footnotesize
\begin{tabular} {l|rrrrrr}
 & Im$\,g_4^{\gamma}$ & Im$\,g_4^Z$ & Im$\,\tilde{\lambda}_{\gamma}$ &
Im$\,\tilde{\lambda}_Z$ & Im$\,\tilde{\kappa}_{\gamma}$ & 
Im$\,\tilde{\kappa}_Z$\\
\hline
&&&&&&\\[-2.5ex]
 no polarisation & 5.1 & 3.6 & 1.8 & 1.4 & 5.6 & 4.2\\
$(P_l^-,P_l^+)=(\mp 80\%,0)$ & 2.3 & 1.8 & 0.84 & 0.68 & 2.7 & 2.1\\
$(P_l^-,P_l^+)=(\mp 80\%,\pm 60\%)$ & 1.4 & 1.1 & 0.48 & 0.39 & 1.6 & 1.3\\
$(P_t^-,P_t^+)=(80\%,60\%)$ & 2.5 & 1.8 & 0.63 & 0.53 & 2.5 & 2.0
\end{tabular}
\normalsize
\end{center}
\end{table}


\begin{table}
\caption{\label{tab:pol3} Same as Table \protect\ref{tab:pol1}, but for
\mbox{$\sqrt{s}=$ 800 GeV}.}
\begin{center}
\leavevmode
\footnotesize
\begin{tabular} {l|rrrrrrrr}
 & Re$\,\Delta g_1^{\gamma}$ & Re$\,\Delta g_1^Z$ & Re$\,\Delta
\kappa_{\gamma}$ & Re$\,\Delta \kappa_Z$ & Re$\,\lambda_{\gamma}$ &
Re$\,\lambda_Z$ & Re$\,g_5^{\gamma}$ & Re$\,g_5^Z$\\
\hline 
&&&&&&&&\\[-2.5ex]
 no polarisation & 4.0 & 3.2 & 0.47 & 0.58 & 1.1 & 0.90 & 3.1 & 2.5\\
$(P_l^-,P_l^+)=(\mp 80\%,0)$ & 1.9 & 1.6 & 0.21 & 0.21 & 0.53 & 0.43 & 
1.6 & 1.3\\
$(P_l^-,P_l^+)=(\mp 80\%,\pm 60\%)$ & 1.1 & 0.97 & 0.14 & 0.13 & 0.29
& 
0.24 & 0.97 & 0.82\\
$(P_t^-,P_t^+)=(80\%,60\%)$ & 1.8 & 1.5 & 0.27 & 0.35 & 0.28 & 0.23 & 
1.7 & 1.3
\end{tabular}
\normalsize
\end{center}
\end{table}


\begin{table}
\caption{\label{tab:pol3imLR} Same as Table \protect\ref{tab:pol1imLR},
but for \mbox{$\sqrt{s}=$ 800 GeV}.}
\begin{center}
\leavevmode
\footnotesize
\begin{tabular} {l|rrrr|rrrr}
 & Im$\,g_1^L$ & Im$\,\kappa_L$ & Im$\,\lambda_L$ & Im$\,g_5^L$ &
$\tilde{h}_-$ & $\tilde{h}_+$ & Im$\,\lambda_R$ & Im$\,g_5^R$\\
\hline  
&&&&&&&&\\[-2.5ex]
no polarisation & 1.5 & 0.74 & 0.18 & 1.5 & 6.0 & --- & 1.2 & 9.0 \\ 
$(P_l^-,P_l^+)=(\mp 80\%,0)$ & 1.5 & 0.60 & 0.17 & 1.3 & 2.4 & --- & 
0.54 & 2.7 \\ 
$(P_l^-,P_l^+)=(\mp 80\%,\pm 60\%)$ & 1.2 & 0.48 & 0.14 & 1.0 & 1.3 & 
--- & 0.29 & 1.4 \\ 
$(P_t^-,P_t^+)=(80\%,60\%)$ & 1.5 & 0.60 & 0.17 & 1.3 & 2.1 & 2.0 &
0.39 & 2.8
\end{tabular}
\normalsize
\end{center}
\end{table}


\begin{table}
\caption{\label{tab:pol4} Same as Table \protect\ref{tab:pol1}, but for
\mbox{$\sqrt{s}=$ 800 GeV} and symmetry (c).}
\begin{center}
\leavevmode
\footnotesize
\begin{tabular} {l|rrrrrr}
 & Re$\,g_4^{\gamma}$ & Re$\,g_4^Z$ & Re$\,\tilde{\lambda}_{\gamma}$ &
Re$\,\tilde{\lambda}_Z$ & Re$\,\tilde{\kappa}_{\gamma}$ & 
Re$\,\tilde{\kappa}_Z$\\
\hline
&&&&&&\\[-2.5ex]
 no polarisation & 4.1 & 3.4 & 1.1 & 0.92 & 4.5 & 3.3\\
$(P_l^-,P_l^+)=(\mp 80\%,0)$ & 2.0 & 1.7 & 0.54 & 0.44 & 2.1 & 1.6\\
$(P_l^-,P_l^+)=(\mp 80\%,\pm 60\%)$ & 1.2 & 1.0 & 0.30 & 0.24 & 1.2 & 1.0\\
$(P_t^-,P_t^+)=(80\%,60\%)$ & 1.8 & 1.6 & 0.28 & 0.23 & 1.9 & 1.5
\end{tabular}
\normalsize
\end{center}
\end{table}


\begin{table}
\caption{\label{tab:pol4im} Same as Table \protect\ref{tab:pol1}, but for
\mbox{$\sqrt{s}=$ 800 GeV} and symmetry~(d).}
\begin{center}
\leavevmode
\footnotesize
\begin{tabular} {l|rrrrrr}
 & Im$\,g_4^{\gamma}$ & Im$\,g_4^Z$ & Im$\,\tilde{\lambda}_{\gamma}$ &
Im$\,\tilde{\lambda}_Z$ & Im$\,\tilde{\kappa}_{\gamma}$ & 
Im$\,\tilde{\kappa}_Z$\\
\hline
&&&&&&\\[-2.5ex]
 no polarisation & 3.8 & 2.8 & 0.72 & 0.60 & 4.0 & 2.9\\
$(P_l^-,P_l^+)=(\mp 80\%,0)$ & 1.6 & 1.3 & 0.34 & 0.28 & 1.8 & 1.4\\
$(P_l^-,P_l^+)=(\mp 80\%,\pm 60\%)$ & 0.93 & 0.79 & 0.19 & 0.16 & 
1.1 & 0.86\\ 
$(P_t^-,P_t^+)=(80\%,60\%)$ & 1.7 & 1.3 & 0.25 & 0.21 & 1.7 & 1.4
\end{tabular}
\normalsize
\end{center}
\end{table}


\begin{table}
\caption{\label{tab:corr} Averages over the absolute values of the
off-diagonal elements in the correlation matrices
(\protect{\ref{eq:wdef}}) in \%, for symmetry~(a) with
$\sqrt{s}=500$~GeV and different beam polarisations.  Apart from the
average over all 28 couplings (last column) we list the averages over
the correlations between L-couplings (LL), between R-couplings (RR)
and those between one L- and one R-coupling (LR).}
\begin{center}
\leavevmode
\begin{tabular} {l|rrrr}
 & LL & RR & LR & all\\
\hline 
&&&&\\[-2.5ex]
 no polarisation & 22 & 42 & 14 & 22\\
$(P_l^-,P_l^+)=(\mp 80\%,0)$ & 22 & 41 & 4 & 16\\
$(P_l^-,P_l^+)=(\mp 80\%,\pm 60\%)$ & 22 & 37 & 2 & 13\\
$(P_t^-,P_t^+)=(80\%,60\%)$ & 20 & 26 & 8 & 15
\end{tabular}
\end{center}
\end{table}

\pagebreak


\appendix

\section{Appendix: Phase conventions of the helicity states}
\label{app-phase}

To make the discrete symmetry properties of the initial state
(cf.~Sect.~\ref{sec-initial}) more apparent, we present in detail our
phase conventions of the helicity states in this appendix.  The
resulting criteria for $CP$ and $R\,CP\tilde{T}$ invariance of the
spin density matrix are shown in Appendix~\ref{app-disc}.  Our
starting point is a Wigner basis of electron and positron states (see
Chapter 16 of \cite{Nachtmann:ta}) defined in the $e^-e^+$~c.m.~system:
\begin{equation}
|e^-({\bf p}, \sigma)\rangle_{\rm W},\;\;\;\;\;\;\;\;\;\; |e^+({\bf
 p}, \sigma)\rangle_{\rm W} \;\;\;\;\;\;\;\;\;\; (\sigma = \pm 1).
\end{equation}
Here ${\bf p}$ is an arbitrary three-momentum in the $e^-e^+$~c.m.,
given in the coordinate system fixed by the \mbox{$e^-e^+ \rightarrow
W^-W^+$} scattering plane (cf.\ Fig.~\ref{fig:pro}), and $\sigma /2$
is the spin component along the positive $z$-axis (we follow the
notation of \cite{Diehl:2002nj} and normalise all spin and helicity
indices to 1).  We set
\begin{equation}
{\bf k}_{\pm} := (0, 0, \pm |{\bf k}|),\;\;\;\;\;\;\;\;\;\;|{\bf k}| =
\frac{1}{2} \sqrt{s - 4 m_e^2}\,,
\end{equation}
where $\sqrt{s}$ is the c.m.~energy of $e^-e^+$.  We define the
helicity states with momentum ${\bf k}_+$ to be the Wigner states
\begin{equation}
|e^{\pm}({\bf k}_+, \tau)\rangle_{\rm H} = |e^{\pm}({\bf k}_+,
 \tau)\rangle_{\rm W}.
\end{equation}
We define the helicity states with momentum ${\bf k}_-$ by a rotation
of $+ \pi$ around the $y$-axis, i.e.\ we set
\begin{equation}
R = \exp (- i \pi J_y),\;\;\;\;\;\;\;\;\;\;|e^{\pm}({\bf k}_-,
 \tau)\rangle_{\rm H} = U(R) |e^{\pm}({\bf k}_+, \tau)\rangle_{\rm H},
\label{eq:defrot}
\end{equation}
where $J_y$ is the angular momentum along $y$.  The transformation
formulae for the Wigner states (see Appendix~J, 16.3 of
\cite{Nachtmann:ta}) give
\begin{equation}
|e^{\pm}({\bf k}_-, \tau)\rangle_{\rm H} = - |e^{\pm}({\bf k}_-,
 \sigma)\rangle_{\rm W} \;\; \epsilon_{\sigma \tau}
\label{eq:defhel}
\end{equation}
with
\begin{equation}
\epsilon = \left( \begin{array}{rr} 0 & \phantom{-}1 \\ -1 & 0 \end{array} \right).
\end{equation}
Here and in the following summation over repeated indices is
understood.  Our sign convention in the exponent of (\ref{eq:defrot})
together with the prescription to rotate around \boldm{\hat{e}_y} by
$+180$ degrees is consistent with the spinors\footnote{Note that,
compared with the coordinate system used here, the spinors in
\protect{\cite{Diehl:2002nj}} are defined in a system rotated by
$\Theta$ around the $y$-axis.}  (103) and (104) of
\cite{Diehl:2002nj}.  For the spin density matrix \boldm{\rho} of the
$e^-e^+$ system in the helicity and Wigner bases we obtain the relation
\begin{eqnarray}
{\phantom{\Big|}}_{\rm H} \Big\langle e^-({\bf k}_+, \tau) e^+({\bf k}_-,
\overline{\tau}) \Big|\, \boldm{\rho} \,\Big| 
e^-({\bf k}_+, \tau') e^+({\bf k}_-,
\overline{\tau}') \Big\rangle_{\rm H} &&
\nonumber\\ 
&& \hspace{-7.5cm} =
{\phantom{\Big|}}_{\rm W} \Big\langle e^-({\bf k}_+, \tau) e^+({\bf k}_-,
\overline{\sigma}) \Big|\, \boldm{\rho} \,\Big| 
e^-({\bf k}_+, \tau') e^+({\bf
k}_-, \overline{\sigma}') \Big\rangle_{\rm W} \;\;
\epsilon_{\overline{\sigma}\, \overline{\tau}} \;
\epsilon_{\overline{\sigma}' \overline{\tau}'}\; ,
\end{eqnarray}
or, in shorthand notation,
\begin{equation}
\rho^{\rm H}_{(\tau \overline{\tau})(\tau' \overline{\tau}')} =
\rho^{\rm W}_{(\tau \overline{\sigma})(\tau' \overline{\sigma}')} \;
\epsilon_{\overline{\sigma}\,\overline{\tau}} \;
\epsilon_{\overline{\sigma}' \overline{\tau}'}\; ,
\end{equation}
where $\rho^{\rm H}$ is the spin density matrix in the helicity basis
and $\rho^{\rm W}$ is the one in the Wigner basis.  The matrix
$\rho^{\rm H}$ is therefore the same as $\rho$ in (\ref{eq:density}).
If the spin density matrix in the Wigner basis factorises, i.e.\ if
\begin{equation}
\rho^{\rm W}_{(\tau \overline{\tau})(\tau' \overline{\tau}')} =
\rho^{\rm W}_{\tau \tau'} \;
\overline{\rho}^{\rm W}_{\overline{\tau}\;\overline{\tau}'}\;,
\end{equation}
it also factorises in the helicity basis, with factors       
\begin{equation}
\rho^{\rm H}_{\tau \tau'} = \rho^{\rm W}_{\tau \tau'}
\;,\;\;\;\;\;\;\;\;\;\;
\overline{\rho}^{\rm H}_{\overline{\tau}\,\overline{\tau}'} =
\overline{\rho}^{\rm W}_{\overline{\sigma}\,\overline{\sigma}'} \;
\epsilon_{\overline{\sigma}\,\overline{\tau}} \;
\epsilon_{\overline{\sigma}' \overline{\tau}'}\;.
\end{equation}
We parameterise $\rho^{\rm W}$ and $\overline{\rho}^{\rm W}$ as usual:
\begin{equation}
\rho^{\rm W}_{\tau \tau'} = \frac{1}{2} \left( \mathbbm1 +
\vec{p}\,^{-} \cdot \vec{\sigma}
\right)_{\tau \tau'}\;,\;\;\;\;\;\;\;\;\;\;
\overline{\rho}^{\rm W}_{\overline{\tau}\,\overline{\tau}'} = \frac{1}{2}
\left( \mathbbm1 + \vec{p}\,^{+} \cdot \vec{\sigma}
\right)_{\overline{\tau}\,\overline{\tau}'}\;,
\end{equation}
where $\vec{p}\,^{\pm}$ are the vectors defined in (\ref{eq:polvec}).
This results in the following form of the spin density matrices in the
helicity basis:
\begin{equation}
\rho^{\rm H}_{\tau \tau'} = \frac{1}{2} \left( \mathbbm1 + \vec{p}\,^{-}
\cdot \vec{\sigma} \right)_{\tau \tau'},\;\;\;\;\;\;\;\;\;\;
\overline{\rho}^{\rm H}_{\overline{\tau}\,\overline{\tau}'} = \frac{1}{2}
\left( \mathbbm1 - \vec{p}\,^{+} \cdot \vec{\sigma}^{\ast}
\right)_{\overline{\tau}\,\overline{\tau}'} ,
\label{eq:densmat}
\end{equation}
as given in (\ref{eq:dens}).


\section{Appendix: \boldm{CP} and \boldm{R\,CP\tilde{T}} invariance of
the initial state}
\label{app-disc}

For a symmetry operation that is defined by a unitary operator $U$
acting on the space of state vectors, invariance of \boldm{\rho} under
this symmetry is expressed as
\begin{equation}
\boldm{\rho} = U^{\dagger} \boldm{\rho}\; U.
\end{equation}
We have to reformulate this matrix equation in component notation in
the helicity basis for the symmetries $CP$ and $R\,CP\tilde{T}$.  The
transformation of the Wigner states under $CP$ is defined by the
unitary operator \cite{Nachtmann:ta}
\begin{equation}
U(CP) |e^{\pm}({\bf p}, \sigma)\rangle_{\rm W} = \mp |e^{\mp}(-{\bf p},
\sigma)\rangle_{\rm W}.
\end{equation}
Hence, for an $e^-e^+$ state in the helicity basis we have
\begin{equation}
U(CP) 
|e^-({\bf k}_+, \tau) e^+({\bf k}_-, \overline{\tau}) \rangle_{\rm H}
= - 
|e^-({\bf k}_+, \overline{\sigma}) e^+({\bf k}_-, \sigma) \rangle_{\rm H}
\;\; \epsilon_{\overline{\sigma}\,\overline{\tau}} \;
 \epsilon_{\sigma \tau}\;,
\end{equation}
where the sign due to the interchange of fermions is taken into
account.  Invariance of \boldm{\rho} under $CP$ then corresponds to:
\begin{equation}
\rho^{\rm H}_{(\tau \overline{\tau})(\tau' \overline{\tau}')} = \rho^{\rm
H}_{(\overline{\sigma} \sigma)(\overline{\sigma}' \sigma')} \; 
\epsilon_{\overline{\sigma}\,
\overline{\tau}} \; \epsilon_{\sigma \tau} \; \epsilon_{\overline{\sigma}'
\overline{\tau}'} \; \epsilon_{\sigma' \tau'}\;,
\label{eq:cpinv}
\end{equation}
which leads to the conditions (\ref{eq:parcond}) on the polarisation
parameters.   

We define the discrete symmetry $\tilde{T}$ by a \emph{unitary}
operator that acts on the Wigner states as follows:
\begin{equation}
U(\tilde{T}) |e^{\pm}({\bf p}, \sigma)\rangle_{\rm W} = - |e^{\pm}(-{\bf p},
\sigma')\rangle_{\rm W} \; \epsilon_{\sigma' \sigma}\;.
\end{equation}
For the combined symmetry \mbox{$U(CP\tilde{T}) = U(CP)U(\tilde{T})$}
we then obtain
\begin{equation}
U(CP \tilde{T}) |e^{\pm}({\bf k}_{\mp}, \tau) \rangle_{\rm H} = \pm
|e^{\mp}({\bf k}_{\mp}, \tau') \rangle_{\rm H} \;
\epsilon_{\tau' \tau}\;.
\end{equation}
Together with a subsequent rotation around the $y$-axis by $+180$
degrees (\ref{eq:defrot}) we have
\begin{equation}
U(R\,CP\tilde{T}) |e^{\pm}({\bf k}_{\mp}, \tau) \rangle_{\rm H} 
= - |e^{\mp}({\bf k}_{\pm}, \tau') \rangle_{\rm H} \; 
\epsilon_{\tau' \tau}\;.
\end{equation}
The transformation of the combined $e^-e^+$ state is then
\begin{equation}
U(R\,CP\tilde{T}) |e^-({\bf k}_+, \tau) e^+({\bf k}_-, \overline{\tau})
\rangle_{\rm H} 
= - |e^-({\bf k}_+, \overline{\sigma}) e^+({\bf k}_-, \sigma)
\rangle_{\rm H} \; \epsilon_{\overline{\sigma}\, \overline{\tau}} \;
\epsilon_{\sigma \tau}\;,
\end{equation}
where again the interchange of two fermions is taken into
account.  Invariance of \boldm{\rho} under $R\,CP\tilde{T}$ then again
leads to (\ref{eq:cpinv}).  So, as for $CP$, {\em full} invariance of
\boldm{\rho} requires (\ref{eq:parcond}), whereas invariance of the
{\em reduced} matrix $\boldm{\tilde{\rho}}$ (\ref{eq:redspindens}) is
trivially fulfilled.


\end{document}